\documentclass{PoS}

\title{Do electromagnetic effects survive in the production of lepton pairs 
in nucleus-nucleus collisions ?}

\ShortTitle{Semicentral collisions}

\author{\speaker{Antoni Szczurek}\thanks{A footnote may follow.}\\
        Institute of Nuclear Physics PAN\\
        E-mail: \email{antoni.szczurek@ifj.edu.pl}}

\abstract{We discuss production of $J/\psi$ and
          $e^+ e^-$ pairs in semicentral and peripheral
          collisions of heavy ions at high energies.
          We focus on photoproduction mechanism.
          We present explanation of results of the ALICE
          collaboration for low transverse momentum production
          of $J/\psi$ as well as results obtained by the STAR
          collaboration for $e^+ e^-$ production, also
          at low transverse momenta.
          We conclude that such effects are important also in this
          new corner (semi-central processes) of the phase space.}

\FullConference{
European Physical Society Conference on High Energy Physics - EPS-HEP2019 -\\
			10-17 July, 2019\\
			Ghent, Belgium}

\begin{document}

\section{Introduction}

Both vector meson and dilepton production are flag
processes in ultraperipheral nucleus-nucleus collisions.
In these processes nuclei do not collide directly one with each
other. Only particles of interest (vector mesons or dileptons)
are produced. Such processes are characterised by small multiplicity.
Those processes are induced by strong electromagnetic fields
sourounding colliding high-energy heavy ions.
Both vector meson production and dilepton production was studied
in traditional nuclear collisions in a broad range of collisions
energies. At high energy collisions one is interested in making a relation
to production of quark-gluon plasma which is a traditional topic of
interest in the field.

Three years ago the ALICE collaboration observed 
$J/\psi$ with very small transverse momenta in peripheral 
and semi-central collisions \cite{ALICE2016}.
This was interpreted in \cite{KS2016} as effect of photoproduction
mechanism which is active also in such a case.

Last year the STAR collaboration observed also enhanced production of 
dielectron pairs with small transverse momenta \cite{Adam:2018tdm}.
Last year we showed \cite{KRSS2019} that this may be interpreted as 
$\gamma \gamma \to e^+ e^-$ processes (with coherent photons) even 
in the semi-central collisions.

In this presentation we briefly review the new processes, which we
call semicentral to be distinguished from central or ultraperipheral
collisions extensively discussed in the literature.

\section{A sketch of the formalism}

We start by showing the general situation in the impact parameter
space.

\begin{figure}[h!]
\centering
\includegraphics[width=4cm]{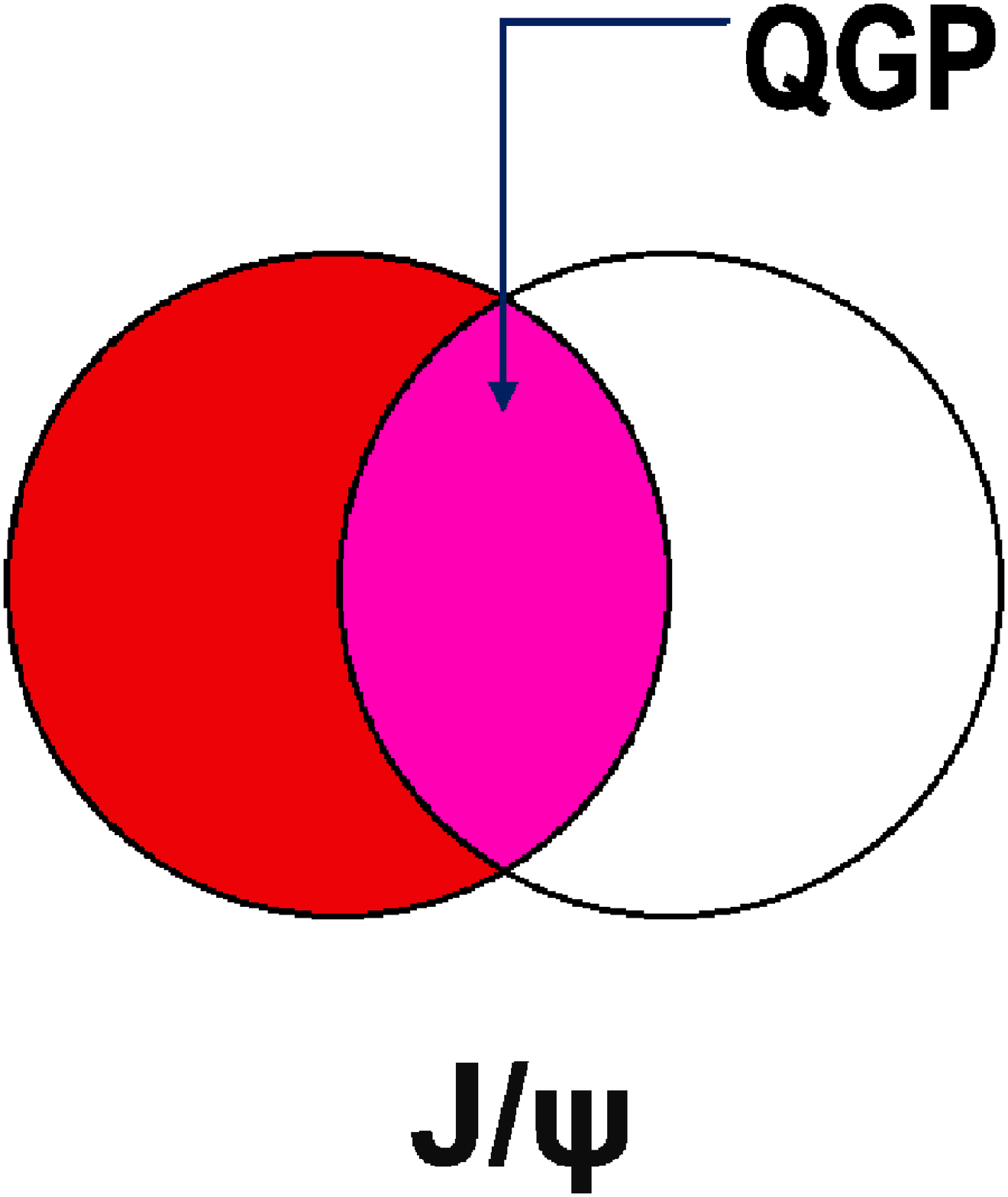}\qquad
\includegraphics[width=4cm]{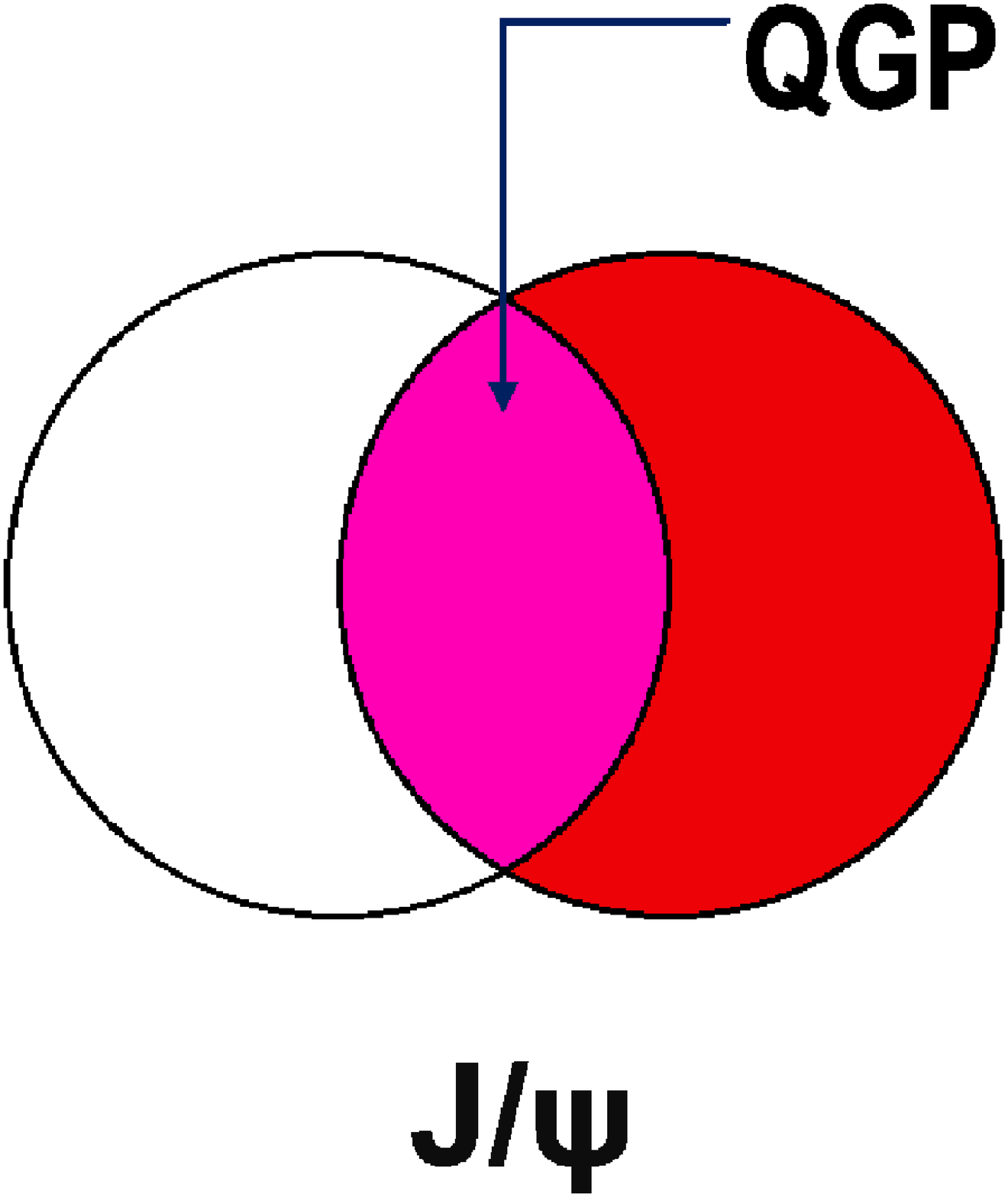}
\caption{Emission of $J/\psi$ - a picture in the plane $x,y$
  perpendicular to the collision axis
    ($z$). In the gray area quark-gluon plasma is created.}
\label{fig:impact_parameter_jpsi}
\end{figure}

In Fig.\ref{fig:impact_parameter_jpsi} we show the situation for
$J/\psi$ production.
The production of meson ``happens'' in the second nucleus.
There are two contributions: (a) either the first (left) or (b) the second
(right) nucleus emits a foton which fluctuates and rescatter in the
partner nucleus. There is a region of overlapping nuclear densities
where quark-gluon plasma is created. Can vector meson ($J/\psi$ in our
case) be produced then ? A comparison with experimental data
may answer this question.

\begin{figure}[h!]
\centering
\includegraphics[width=5cm]{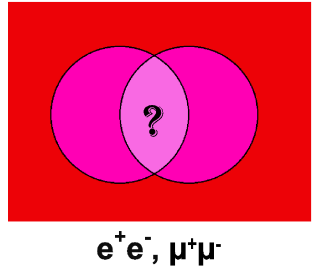}
\caption{Dilepton production - a picture in the plane $x,y$
  perpendicular to the collision axis ($z$).
The area with the question mark is the region where quark-gluon plasma
is created.
}
\label{fig:impact_parameter_epem}
\end{figure}

A $b$-space picture for dileption production is represented in 
Fig.\ref{fig:impact_parameter_epem}. In general, the dilepton pair
may be produced everywhere in the $(b_x, b_y)$ space.
It is worth of noting that a big contribution may come from the red
region, i.e. outside of the nuclei. Other regions have less
sure status.

The main ingredient for the photon-photon fusion mechanism is 
the flux of photons for an ion of charge $Z$ moving
along $z$-axis at a given impact parameter with the relativistic 
parameter $\gamma$. 
With the nuclear charge form factor $F_{\rm em}$ as an input the flux
can be calculated as~\cite{Bertulani:1987tz, Baur:2001jj}
\begin{eqnarray}
N(\omega,b) 
&&= {Z^2 \alpha_{\rm EM} \over \pi^2} 
\Big| \int_0^\infty  dq_t {q_t^2   F_{\rm em}(q_t^2 + {\omega^2 \over \gamma^2} )   
	\over q_t^2 + {\omega^2 \over \gamma^2} } J_1(b q_t) \Big|^2\, ,  
\label{eq:WW-flux}
\end{eqnarray}
where $J_1$ is a Bessel function, $q_t$ is photon transverse momentum
and $\omega$ is photon energy. 
We calculate the form factor from the Fourier transform of the 
nuclear charge density.

The differential cross section for dilepton ($l^+ l^-$) production via 
$\gamma \gamma$ fusion at fixed impact parameter of a nucleus
nucleus collision can then be written as
\begin{eqnarray}
{d \sigma_{ll} \over d\xi d^2 b } =  
\int d^2b_1 d^2b_2 \, \delta^{(2)}(\vec{b} - \vec{b}_1 - \vec{b}_2) N(\omega_1,b_1) N(\omega_2,b_2) 
{d \sigma(\gamma \gamma \to l^+ l^-; \hat s) \over d p_t^2} \ ,
\end{eqnarray}
where the phase space element is $d\xi = dy_+ dy_- dp_t^2$ with $y_\pm$, $p_t$ and $m_l$ the single-lepton 
rapidities, transverse momentum and mass, respectively,
and
\begin{eqnarray}
\omega_1 = {\sqrt{p_t^2 + m_l^2} \over 2} \, ( e^{y_+} + e^{y_-} ) \  , \  
\omega_2 = {\sqrt{p_t^2 + m_l^2} \over 2} \, ( e^{-y_+} + e^{-y_-} ) \ , \ \hat{s} = 4 \omega_1 \omega_2 \ .
\end{eqnarray}
As can be seen from Eq.(\ref{eq:WW-flux}), the transverse momenta
of the photons have been 
integrated out, and in this approximation dileptons are produced 
back-to-back in the transverse plane. 

An exact calculation of the pair-$P_T$ dependence is, in general,  
rather involved.
In Ref.\cite{KRSS2019} we performed a simplified calculation
using $b$-integrated transverse momentum dependent photon fluxes.

\section{Selected results}

\subsection{$J/\psi$ production}

\begin{figure}[!h]
\centering
\includegraphics[scale=0.35]{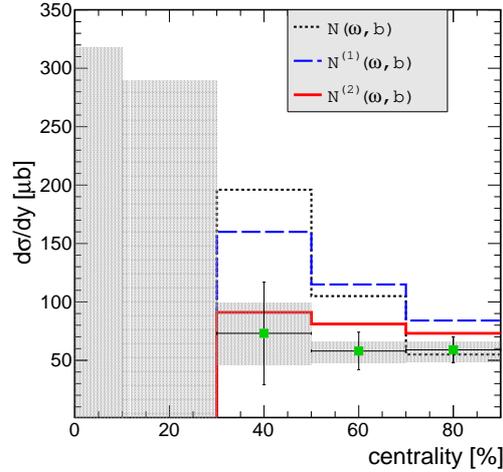}
\caption{$\Delta \sigma / \Delta y$ cross sections for different 
centrality bins. Theoretical results for different
models of the photon flux are compared with the
ALICE data \cite{ALICE2016}.
The shaded area represents the experimental uncertainties.}
\label{fig:flux_hist}
\end{figure}

In Fig.\ref{fig:flux_hist} we show our results (cross section in the
ALICE rapidity interval) starting from
centralities bigger than $30 \%$.
The ALICE Collaboration could not extract actual values
of the cross section for the two lowest centrality bins. 
The results when using standard photon fluxes exceed the ALICE data.
Rather good agreement with the data is achieved 
when the overlap region in Fig.\ref{fig:impact_parameter_jpsi} 
is removed (which corresponds to $N^{(2)}$ in \cite{KS2016}).

The $J/\psi$ corresponding to the photoproduction mechanism are
concentrated in the region of very small transverse momenta.
So far, according to my knowledge, there were no attempts to describe 
transverse momentum distributions of $J/\psi$ mesons.
In the $b$-space approach it is just the Dirac delta.

\subsection{$e^+ e^-$ production}

\begin{figure}[!t]
	\includegraphics[scale=0.35]{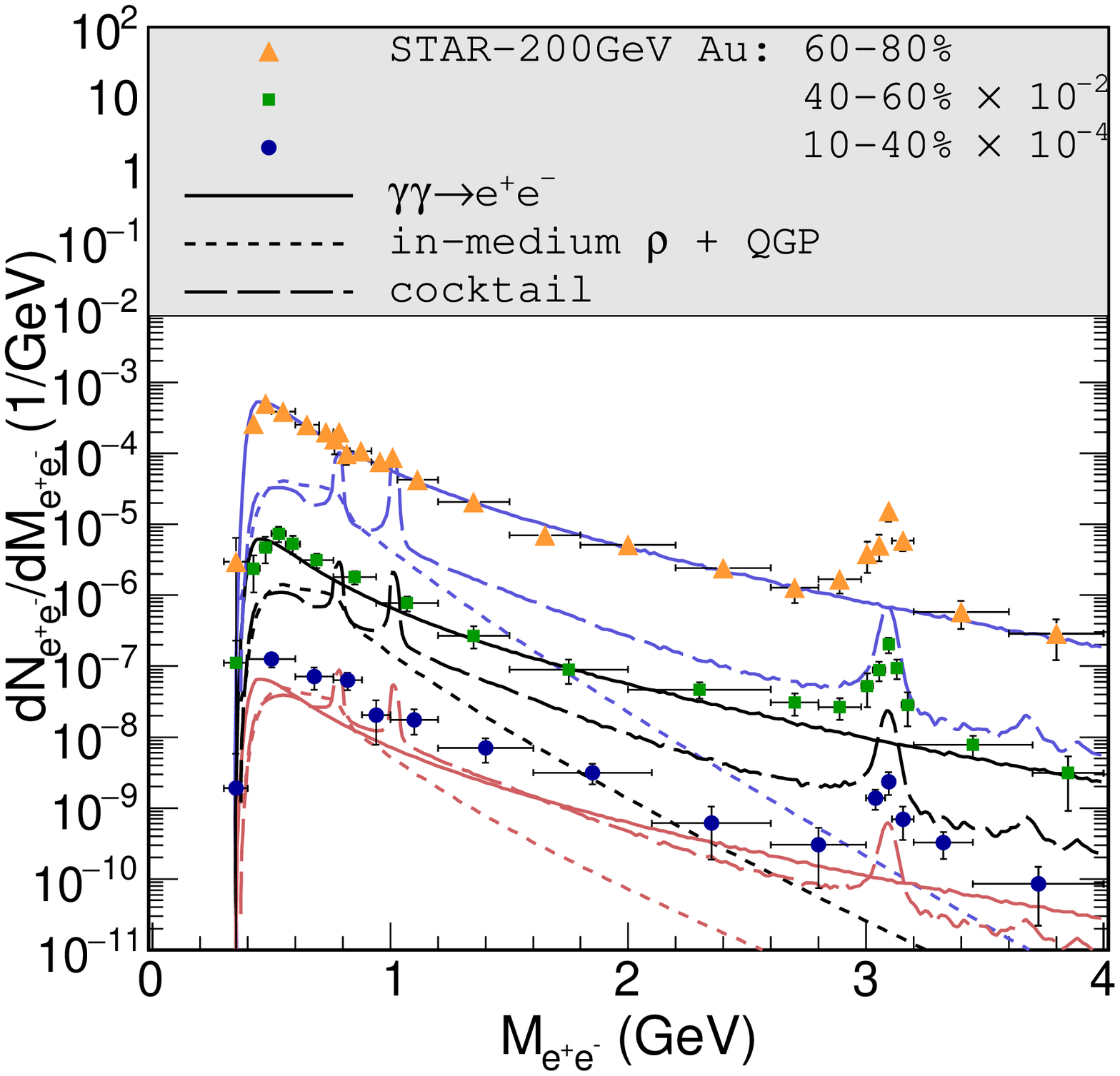}
	\includegraphics[scale=0.35]{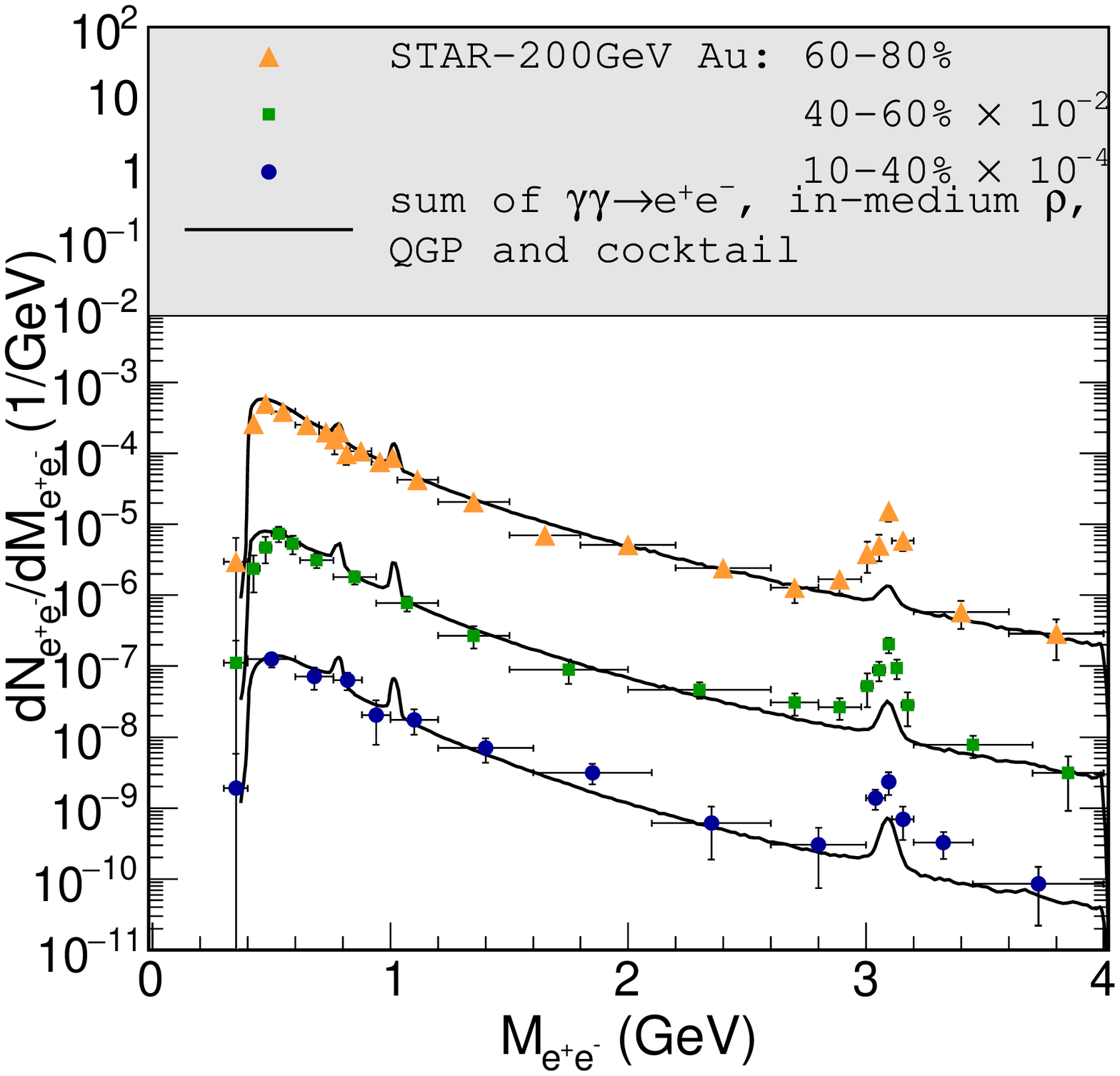}
	\caption{
		Left: Dielectron invariant-mass spectra for pair-$P_T$$<$0.15\,GeV in Au+Au($\sqrt{s_{NN}}$=200\,GeV)
		collisions for 3 centrality classes including experimental acceptance cuts ($p_t>$0.2\,GeV, 
		$|\eta_e|$$<$1 and $|y_{e^+e^-}|$$<$1) for $\gamma \gamma$ fusion (solid lines), thermal radiation 
		(dotted lines) and the hadronic cocktail (dashed lines);
		right panel: comparison of the total sum (solid lines)
                to 
the STAR data~\cite{Adam:2018tdm}.
	}
	\label{fig:STAR_Mll}
\end{figure}

The dielectron invariant-mass distribution is shown
in Fig.\ref{fig:STAR_Mll}. Here we show results for
$P_{T} <$ 0.15~GeV (transverse momentum of the pair) and three different 
centrality classes as selected in the STAR analysis: peripheral (60-80\%), 
semi-peripheral (40-60\%) and semi-central (10-40\%) collisions. 
In peripheral collisions the photon-photon contribution dominates while
in semi-central collisions all three contributions are of similar magnitude. 
Their sum is in good agreement with the STAR data, except for 
the $J/\psi$ peak region. Our calculations contain only incoherent $J/\psi$ 
production, from binary nucleon-nucleon collisions.
We conjecture that the missing contribution is due to a coherent 
contribution~\cite{KS2016}.

\begin{figure}[!t]
\centering
	\includegraphics[scale=0.35]{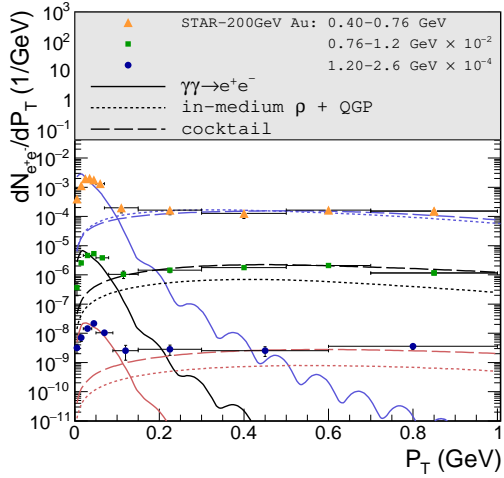}
	\caption{
		$P_T$ spectra of the individual contributions (line styles as in the previous figure)
		in 3 different mass bins for the 60-80\% centrality of the 
		Au+Au collisions ($\sqrt{s_{NN}}$=200\,GeV), compared to
                the STAR data~\cite{Adam:2018tdm}.
	}
	\label{fig:STAR_pt}
\end{figure}

Can we describe also transverse momentum distributions of 
the $e^+ e^-$ pair ? The $b$-space approach sketched above is
not enough to calculate transverse momentum distributions of the pair
which is a Dirac delta function in the approximation made there.
To calculate the distributions in transverse momentum of the pair
one needs rather momentum approach. In our recent paper we considered
a $k_t$-factorization approach. In this approach one considers fluxes
of photons which depend on photon energies and photon transverse
momenta. In this way one gets rather shapes of the transverse momentum 
distributions. The normalization in a given bin of centrality may be
taken from the $b$-space calculation. Our results are shown in 
Fig.\ref{fig:STAR_pt}. We show also contributions of other processes
considered in \cite{KRSS2019}.

In general, there could be also dependence of transverse
momentum distribution of the pair on impact parameter (centrality
of the collision). In our opinion there is no correct formalism
in the literature in this respect. We will discuss this issue elsewhere.

\section{Conclusions}

Here I have briefly reviewed a new category of processes discovered
in last three years. I have started my presentation from
low transverse momentum production of $J/\psi$. It was shown that
cross section for such processes strongly depends on the region
of impact parameter space. In particular, the region (in impact
parameter space) of overlapping
nuclear densities is interesting as there quark-gluon plasma is
created. We have discussed a potential effect of the plasma on
production of $J/\psi$. By comparing to the ALICE data we are
inclined to conclude that the production of $J/\psi$ in this region
seems to be suppressed. This provides a new information on the space-time
picture of the vector meson creation.

I have presented also results for dielectron production for RHIC
energies. The low transverse momentum dielectron production can be
understood as due to $\gamma \gamma \to e^+ e^-$ subrocess in
heavy ion collisions. We obtained a good agreement with the STAR data
for different centralities. I have pointed out the essential geometrical
difference between $e^+ e^-$ and $J/\psi$ production.
In our recent paper we presented also first trial to understand
transverse momentum distributions of $e^+ e^-$ pairs.
This is a very interesting observable which requires further attention
in the future.

\vspace{0.5cm}

{\bf Acknowledgement}

I am indebted to Mariola K{\l}usek-Gawenda, Ralf Rapp and Wolfgang
Sch\"afer for collaboration on the issues presented here.

\end{document}